\documentclass[preprint]{aastex}
\usepackage{extarrows}
\usepackage[breaklinks,hyperindex]{hyperref}
\def\be{\begin{equation}}
\def\ee{\end{equation}}
\newcommand{\etal} {{\it et~al.\ }}

\slugcomment{{\sc Accepted by ApJL: January 27, 2009} } 
\begin{document}

\shorttitle{Constraints on MSP Evolution}
\shortauthors{K{\i}z{\i}ltan  \& Thorsett}
\title{Constraints on Pulsar Evolution: The Joint Period-Spindown Distribution of Millisecond Pulsars}
\author{B\"ulent K{\i}z{\i}ltan \& Stephen E. Thorsett}
\affil{Department of Astronomy \& Astrophysics, University of California, Santa Cruz, CA 95064; bulent@astro.ucsc.edu
}
\begin{abstract}

We calculate the joint period-spindown ($P-\dot{P}$) distributions of millisecond radio pulsars (MSRP) for the standard evolutionary model in order to test whether the observed MSRPs are the unequivocal descendants of millisecond X-ray pulsars (MSXP). The $P-\dot{P}$ densities implied by the standard evolutionary model compared with observations suggest that there is a statistically significant overabundance of young/high magnetic field MSRPs. Taking biases due to observational selection effects into account, it  is unlikely that MSRPs have evolved from a single coherent progenitor population that loses energy via magnetic dipole radiation after the onset of radio emission. By producing the $P-\dot{P}$ probability map,  we show with more than 95\% confidence that the fastest spinning millisecond pulsars with high magnetic fields, e.g. PSR B1937+21, cannot be produced by the observed MSXPs within the framework of the standard model.

\end{abstract}


	\section{\label{sec:intro}Introduction}

Millisecond pulsars are commonly believed to be descendants of normal neutron stars that have been spun-up and recycled back as radio pulsars by acquiring angular momentum from their companion during the low-mass X-ray binary (LMXB) phase \citep{ACR82,RS82}.

There are about $\sim$20 high confidence nuclear or accretion powered (see Table~1) millisecond X-ray pulsars (MSXPs) which are thought to be the progenitors of millisecond radio pulsars (MSRPs) \citep{WK98}. These MSXPs may become observable in radio wavelengths once accretion ceases, or the column density of the plasma from the fossil disk around the neutron star becomes thin enough to allow vacuum gap formation that leads to the production of coherent radio emission. Towards the end of the secular LMXB evolution, as accretion rates fall below a critical value above which detection presumably may be hampered due to absorption or dispersion \citep{TBE94}, the neutron star can re-appear as a MSRP.

Although the connection between LMXBs and MSRPs has been significantly strengthened after the discovery of quasi-periodic kHz oscillations and X-ray pulsations in some transient X-ray sources \citep{WK98, MSS02, GCM02, GMM05}, no radio pulsations from MSXPs have been detected so far \citep{BBP03}. 

At the end of the recycling process the neutron star will reach an equilibrium period \citep{BH91} which is approximated by the Keplerian orbital period at the Alfven radius \citep{GL92}:

\be
P_{eq} \sim 1.9\, ms\, B_{9}^{6/7} \biggl (\frac{M}{1.4 \,M_{\sun}}\biggr)^{-5/7} \biggl(\frac{\dot{m}}{\dot{M}_{Edd}}\biggr)^{-3/7} R^{16/7}_{6}
\ee
where $B_{9}$ and $R_{6}$ are the neutron star surface magnetic dipole field and radius in units of $10^{9}$ G and $10^{6}$ cm respectively. The Eddington limited accretion rate $\dot{M}_{Edd}$ for a neutron star typically is $\sim10^{-8}\,M_{\sun}\,yr^{-1}$ above which the radiation pressure generated by accretion will stop the accretion flow. This equilibrium period combined with the dominant mechanism for energy loss delineates the subsequent kinematics of the spun-up millisecond pulsar. The magnetic dipole model then implies a ``spin-up region'' ($\dot{P}\, \sim\, P_{0}^{4/3} $) \citep[see][]{ACW99} on which the recycled neutron stars will be reborn as MSRPs. At the end of the active phase, MSXPs accreting with $\dot{m}$ and  spinning with $P_{eq}$, presumably transition into a MSRPs with an initial spin period of $P_{0} \sim P_{eq}$.

In the standard spin-down model, the MSRP evolution is driven by pure magnetic dipole radiation, i.e. braking index $n=3$ in vacuum \citep[see][]{MT77, LK04}.  Alternative energy loss mechanisms such as multipole radiation or gravitational wave emission, especially during the initial phases of the reborn millisecond pulsars, have been suggested by several authors  \citep{K91,B98} but have yet to be observationally corroborated. Advanced Laser Interferometer Gravitational Wave Observatory (LIGO) will be able to probe the frequency space at which millisecond pulsars are expected to radiate gravitational waves, thereby putting stringent constraints on the micro physics of millisecond pulsars.

The advances in radio observations, increased sky coverage with deep exposures of current surveys combined with robust post-bayesian statistical techniques that incorporate minimal assumptions, give us unprecedented predictive power on the joint period-spindown ($P-\dot{P}$) and implied magnetic field ($B$) distributions.

In this {\it Letter}, we attempt to go beyond phenomenological arguments and test whether MSXPs can produce the characteristics of the observed MSRPs within the framework of the standard model \citep[and references therein]{BH91}.
	
	\section{\label{sec:dist}The Joint Period - Spindown ($P-\dot{P}$) Distribution}
							
The evolution of millisecond pulsars can be consistently described in terms of {\bf i)} the equilibrium period distribution ($D$) of MSXPs at the end of the LMXB evolution {\bf ii)} the mass accretion rates ($\dot{M}$) of the progenitor population during the recycling process  {\bf iii)} Galactic birth rates ($R$), and {\bf iv)} the dominant energy loss mechanism after the onset of radio emission.  

			\subsection{\label{sec:stat}Statistics}

We devise a semi-analytical evolution function $\mathcal{E}$ to parametrize the evolution of millisecond pulsars after the accretion phase, which can be described in closed form as:
\begin{eqnarray}
\displaystyle\sum_{i=0}^{r} \mathcal{E}(D_{i},\dot{M}_{i},R_{i}\, | \, \alpha_{i}^{k},\beta_{i}^{k})  \xrightarrow{n=3} \mathcal{PDF}(P,\dot{P})  \label{stat.eq} 
\end{eqnarray}
where $\mathcal{PDF}$ is the probability distribution function. The shape parameters $\alpha$ and $\beta$ define the distributions (i.e., $D,\dot{M}, R$ for k=1,2,3) for the Beta functions\footnote{Beta functions are commonly preferred in Bayesian statistics as the least restrictive and most flexible prior distributions. It can take the form of an uninformative (e.g. uniform) prior, a monotonic line, concave, convex, unimodal (e.g. normal) or any extreme combinations of these shapes.}\citep{EHP00} inferred from observations at each Monte-Carlo realization ``r''. 

The evolution function $\mathcal{E}$ is built by randomly choosing initiation seeds from a period distribution $D$, which is then convolved via the standard model to consequently sample the $P-\dot{P}$ parameter space. For the observed MSXPs, the period distribution which seeds will be randomly chosen from is the observed $P_{MSXP}$ distribution (table~1). We uniquely construct a ``relaxed multidimensional Kolmogorov-Smirnov (K-S) filter'' (fig.~\ref{fig:probdist}) to check population consistencies by calculating the 2D K-S \citep{FF87} probabilities ($P_{2DK-S}$) between observed MSRPs and the synthetic population that is formed by these properly evolved progenitor seeds. The filtering is reiterated for each realization to obtain synthetic populations with consistent distributions as:
\begin{eqnarray}
D\,(\alpha_{i}^{1},\beta_{i}^{1})  \xrightarrow{filter}  D_{i} \label{statD.eq} \\
\dot{M}(\alpha_{i}^{2},\beta_{i}^{2})   \xrightarrow{filter}  \dot{M}_{i}   \label{statM.eq} \\
R\,(\alpha_{i}^{3},\beta_{i}^{3})  \xrightarrow{filter}  R_{i}  \label{statR.eq}
\end{eqnarray}
which is then used to construct the $\mathcal{PDF}$ in Equation ~\protect\ref{stat.eq}.

Nominally any $P_{2DK-S} > 0.2$ value would imply consistent populations in a 2D K-S test. By allowing $0.005 < P_{2DK-S} < 0.2$ with lower fractions (see fig.~\ref{fig:probdist}), we oversample outliers to compensate for possible statistical fluctuations and contaminations. A peak sampling rate around the nominal acceptance value of $P_{2DK-S}\sim0.2$ is the most optimal scheme that prevents strong biases due to over or under-sampling. The main goal for oversampling outliers and relaxing the K-S filter is to test whether the standard model can at least marginally produce very fast millisecond pulsars with relatively high magnetic fields like PSR B1937+21.

The predictive significance of the $P-\dot{P}$  distribution for the probability map (fig~\protect\ref{fig:MSPs}) is obtained from a Monte-Carlo run with $r=10^{7}$ valid realizations that produce consistent synthetic samples. Whilst sampling the $P-\dot{P}$ space, no assumptions were made regarding the progenitor period distribution ($D$), the accretion ($\dot{M}$), or the Galactic birth ($R$) rates. The filter  (eq.~\protect\ref{statD.eq}, \protect\ref{statM.eq}, \protect\ref{statR.eq}) is implicitly driven by the observed MSRPs.
	
Fig \protect\ref{fig:MSPs} shows the expected $P-\dot{P}$ distribution for the standard model assuming that MSRPs have evolved from a progenitor population similar to the observed MSXPs. We do not include MSRPs in globular clusters because the $P-\dot{P}$ values in these cases may not necessarily be the sole imprint of the binary evolution, but can be significantly changed by possible gravitational interactions due to the crowded field. To explore the extend of the effects of an unevenly sampled progenitor population, we also show the region in the $P-\dot{P}$ space that is sensitive to alternative $P_{MSXP}$ distributions. The probability map is overlaid with the observed MSRPs. 

	\section{\label{sec:dis}Discussion and Conclusions}

The discovery of millisecond pulsations from neutron stars in LMXBs has substantiated the theoretical prediction that links MSRPs and LMXBs. Since then, the recycling process that produces MSRPs on a spin-up region from LMXBs, followed by spin-down due to dipole radiation has been conceived as the ``standard evolution'' of millisecond pulsars. However, the question whether all observed MSRPs could be produced within this framework has not been quantitatively addressed until now.

The standard evolutionary process produces millisecond pulsars with periods ($P$) and spin-downs ($\dot{P}$) that are not entirely independent. The possible $P-\dot{P}$ values that MSRPs can attain are {\it jointly} constrained by the equilibrium period distribution ($D$) of the progenitor population, the mass accretion rates ($\dot{M}$) during the recycling process and the dominant energy loss mechanism after the onset of radio emission. 

\begin{figure} [t!]
	\includegraphics[width=4.8in,angle=90]{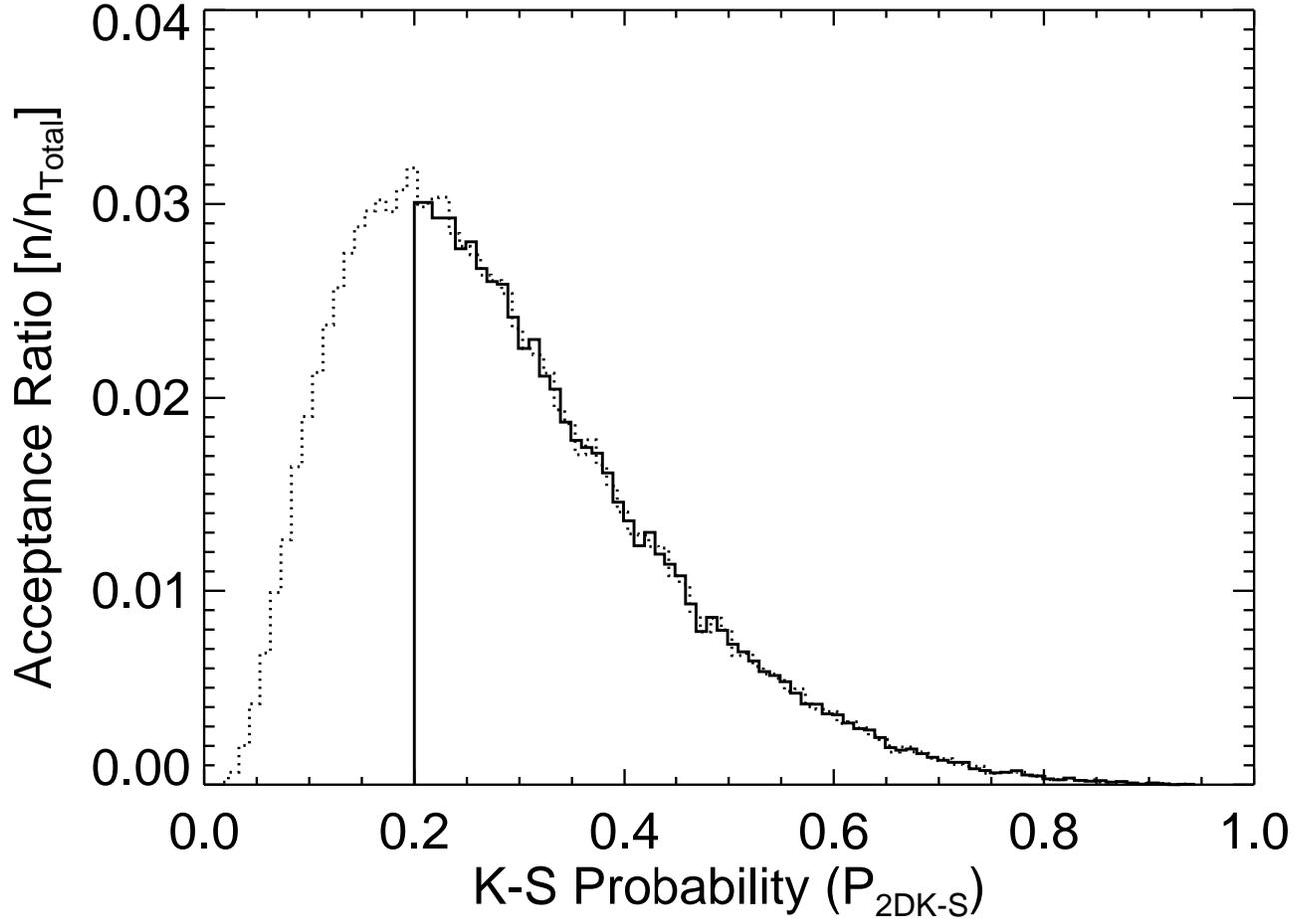}
	\figcaption{\small{The K-S probability distribution of the synthetic populations used to sample the joint $P-\dot{P}$ parameter space. The relaxed 2D K-S filter allows additional acceptance of populations with $0.05 < P_{2DK-S} < 0.2$ by oversampling the extreme outliers of the $P-\dot{P}$ distribution in order to probe for possible contaminations and extreme fluctuations. The distribution also shows how optimally the $P-\dot{P}$ parameter space is sampled with a peak sampling rate around the nominal acceptance value of $ P_{2DK-S}\sim0.2$. The dotted line is the relaxed 2D K-S filter for the synthetic populations that is used to construct the predictive distribution in fig~\ref{fig:MSPs}. The solid line is the conventional K-S filter that would only accept strictly consistent $P-\dot{P}$ samples. }
	\label{fig:probdist}    
	 }	
	\end{figure}

\begin{figure*}[t]
	\includegraphics[width=4.8in,angle=90]{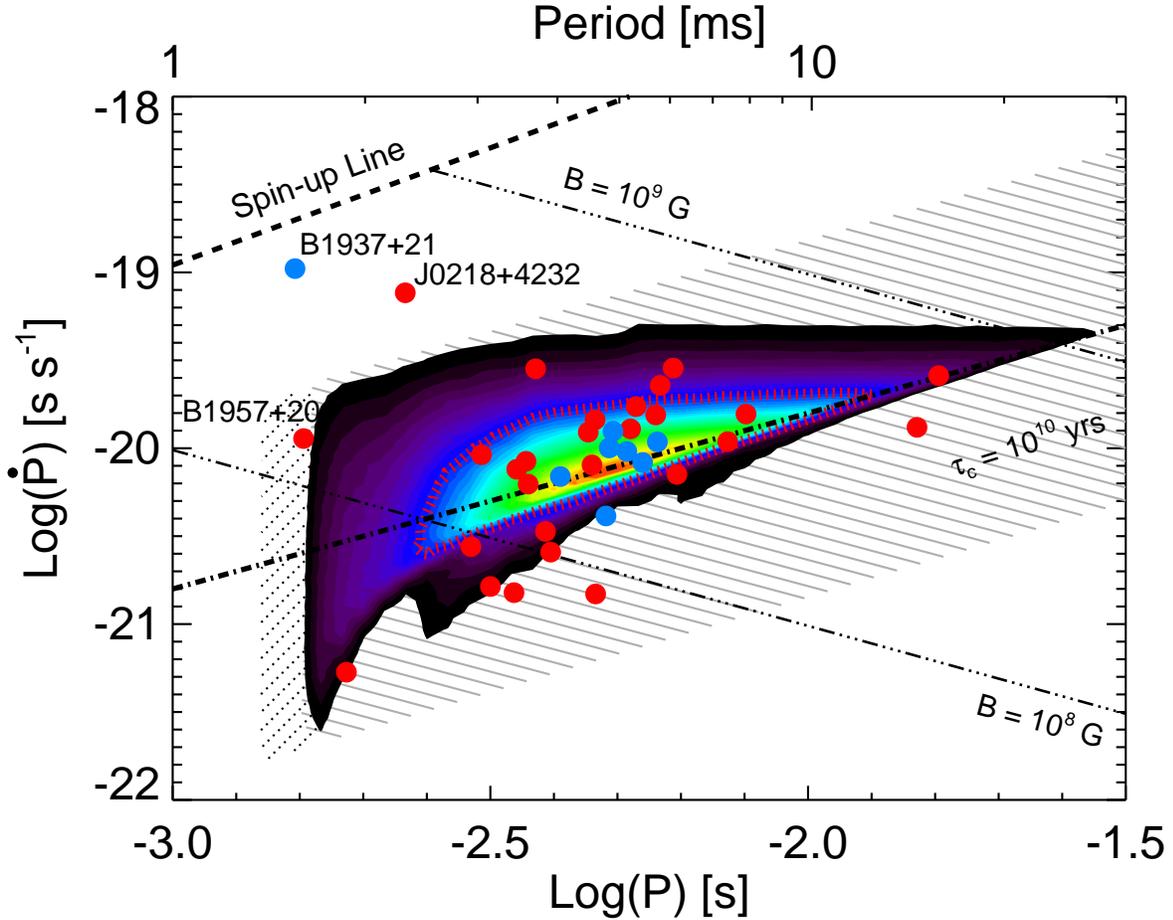}
	\figcaption{\small{The $P-\dot{P}$ distribution of millisecond pulsars for the standard model vs. observed MSRPs. The color contours are the expected MSRP distribution for the observed MSXPs, with red representing the highest density. Any MSRP outside of the color contours cannot be produced by the observed MSXPs with more than 95\% confidence.The dashed red line is the 68\%  confidence limit of the expected $P-\dot{P}$ distribution taking the observed MSXPs as the progenitor population. MSRPs within the shaded region may be produced by MSXPs with spin distributions different than what is observed. The area shaded with lines assumes a maximum spin frequency $\nu_{max}=619$ Hz, which the fastest spinning observed MSXP (i.e. 4U~1608$-$52). The shaded area extends to the dotted region if the maximum spin frequency for MSXPs is allowed to be $\nu_{max}=760$ Hz which is the theoretical upper limit predicted by \protect\cite{CMM03}. The blue and red filled circles are the observed MSRPs in single and binary systems respectively. The area outside of the shaded region is not sensitive to the prior, i.e. the observed MSRPs outside of the shaded area cannot be produced consistently by the standard model for any MSXP spin distribution with more than 95 \% confidence. The spin-down values for the observed MSRPs are corrected for secular acceleration \protect\citep{S70, CTK94}. The spin-up line ($\dot{P}\sim \dot{m} P_{0}^{4/3}$) for $\dot{m}=\dot{M}_{Edd}$ and the characteristic age line for $\tau_{c}=10^{10}$ yrs are shown with dashed and dash-dotted lines.}
	\label{fig:MSPs}   
   }	
	\end{figure*}

In order to test whether the observed MSRPs can be reconciled with a single coherent progenitor population that evolves via magnetic dipole braking after the spin-up process, we have produced the predictive joint $P-\dot{P}$ distribution of MSRPs for the standard model. We did not put restrictions on any of the parameters that drive the evolution. Acceptable $D,\dot{M}$ and $R$ values were implicitly filtered.  We have relaxed the K-S filter (see fig.~\ref{fig:probdist}) in order to oversample outliers and see whether it is even remotely feasible to produce young millisecond pulsars, like PSR B1937+21 or J0218+4232, that have higher B fields. The color contours in Figure~\ref{fig:MSPs} represent the $P-\dot{P}$ densities for MSRPs that are direct descendants of observed MSXPs (i.e. initial spin periods $P_{0}\sim P_{MSXP}$). 

The standard evolutionary model is able to successfully produce the general demographics of older MSRPs. It fails, however, to predict the younger and fastest MSRP sub-population that have higher $B$ fields.

Accretion rates that MSRPs have experienced during their accretion phase deduced from observed $P-\dot{P}$ values, combined with the observed MSXP period distribution ($D \equiv  P_{MSXP}$) produces mostly older MSRPs, including MSRPs with spin-down ages $\tau_{c} > 10^{10}$ yrs. Figure ~\ref{fig:MSPs} shows clearly that the apparent enigma of millisecond pulsars with spin-down ages older than the age of the Galaxy is mainly a manifestation of very low accretion rates during the late stages of the LMXB evolution. 

On the other hand, no physically motivated $P_{MSXP}$ distribution has been able to produce the whole MSRP population consistently. The observed period distribution of MSXPs is likely to be under-sampled due to observational selection effects. It is also possible that some neutron stars in LMXBs simply do not produce observable pulses. In order to understand how the predicted $P-\dot{P}$ distribution is affected by different MSXP period distributions, we have estimated the whole extend of the $P-\dot{P}$ region that is sensitive to the prior. The values that may be produced for different $P_{MSXP}$ distributions are shown by the shaded areas in Figure ~\ref{fig:MSPs}. No MSXP period distribution could mimic the observed relative ratios of young/old pulsars with high B fields. The fraction of the observed young/old MSRPs with high $B$ fields is higher than what the standard model predicts by several orders of magnitude. This may further be exacerbated by strong selection effects that limit our ability to observe very fast millisecond pulsars \citep{HRS07}. The choice of a standard K-S test instead of the relaxed 2D K-S only increases the statistical significance. Hence, we argue that young millisecond pulsars with higher magnetic fields (e.g. PSR B1937+21) are inconsistent with the standard model. 

Therefore, it is tempting to suggest that the fastest spinning millisecond pulsars, in particular PSR B1937+21, may originate from a different evolutionary channel.  While it appears that ordinary magnetic-dipole spin down from a source population similar to the observed MSXPs is adequate to explain the great majority of observed MSRPs, the low final accretion rates that are required cannot be reconciled with the high accretion rates needed to produce the fastest, youngest pulsars. We believe that it is necessary to posit the existence of a separate class of progenitors, most likely with a different distribution of magnetic fields, accretion rates and equilibrium spin periods, presumably among the LMXBs that have not been revealed as MSXPs.  Understanding this additional channel is clearly critical to developing a natural solution to the long lasting ``birth rate problem'' \citep[see, e.g.][]{KN88}.

It is also possible that the standard evolutionary model fails at another point. For example, if MSRPs during some portion of their evolution lose energy through a dominant mechanism other than magnetic dipole radiation (e.g. multipole radiation, gravitational wave or neutrino emission), then the evolution of pulsars through the $P-\dot{P}$ diagram could be complex. 

A combination of the above mentioned factors (i.e. alternative progenitors and subsequent non-standard radiation) are then likely to play a role in millisecond pulsar evolution. A MSXP period distribution that has sharp multimodal features coupled with non-standard energy loss mechanisms may be able to reconcile for the joint $P-\dot{P}$ distribution of millisecond pulsars.

\acknowledgements 
The research presented here has made use of the August 2008 version of the ATNF Pulsar Catalogue \citep{MHT93}. The authors acknowledge NSF grant AST-0506453.


\begin{deluxetable}{lll}
\label{tab:lmxb}
\tablecolumns{2}
\tabletypesize{\normalsize}
\singlespace
\tablewidth{0pt}
\tablecaption{Accretion and nuclear powered pulsars}
\tablehead{ 
\colhead{$\nu_{spin}$ [Hz]} & 
\colhead{Pulsar }  &
\colhead{Reference} 
}
\startdata
619\    &\hbox{4U~1608$-$52}   		& 	\cite{HCG03}		 \\
601\    &\hbox{SAX~J1750.8$-$2900}	 &	 \cite{KZH02}		\\
598\    &\hbox{IGR J00291$+$5934} 	& 	\cite{MSS04} \\  
589\    &\hbox{X~1743$-$29} 		          &	 \cite{SJG97}		 \\
581\    &\hbox{4U~1636$-$53}  		&	\cite{ZLS97}	   	\\
567\    &\hbox{X~1658$-$298}  		 &	\cite{WSF01}	         \\    
549\    &\hbox{Aql~X--1}   	 			&	\cite{ZJK98}	         \\
524\    &\hbox{KS~1731$-$260}  		&	\cite{SMB97}	         \\
435\    &\hbox{XTE~J1751$-$305} 		&	\cite{MSS02}	         \\	    
410\    &\hbox{SAX~J1748.9$-$2021}   	&     	\cite{KZH03}	         \\
401\    &\hbox{SAX~J1808.4$-$3658} 	& 	\cite{WK98}		\\   
& &									 \cite{CM98} 		 \\
377\    & \hbox{HETE~J1900.1$-$2455}	&	  \cite{KMV06}		 \\
363\    &\hbox{4U~1728$-$34}			& 	\cite{SZS96}		  \\
330\    &\hbox{4U~1702$-$429}  		&	\cite{MSS99}		  \\ 		    
314\    &\hbox{XTE~J1814$-$338}		& 	\cite{MS03}		  \\	    
270\    &\hbox{4U~1916$-$05}			&	\cite{GCM01}		  \\
191\    &\hbox{XTE~J1807.4$-$294}		&	\cite{MSS03}  		  \\ 	    
185\    &\hbox{XTE~J0929$-$314}  		&	\cite{GCM02}	       	   \\    
45\     &\hbox{EXO 0748$-$676}   		&	\cite{VS04} 		    \\   
\hline
\enddata
\tablecomments{The millisecond pulsar progenitor seeds used to construct the cumulative synthetic MSRP population for the observed $P_{MSXP}$.} 
\end{deluxetable}


\end{document}